%% file: main.tex
\documentclass[11pt]{article}
\usepackage{customstyle}
\usepackage{hyperref}
\usepackage{titling}
\usepackage{amsfonts}
\usepackage{float}

\title{Securing Unbounded Differential Privacy Against Timing Attacks}
\author{Zachary Ratliff\thanks{Harvard University \& OpenDP. Email: \texttt{zacharyratliff@g.harvard.edu}. Supported in part by Cooperative Agreement CB20ADR0160001 with the Census Bureau, and in
part by Salil Vadhan’s Simons Investigator Award}
\and
Salil Vadhan\thanks{Harvard University \& OpenDP. Email: \texttt{salil\_vadhan@harvard.edu}}
}
\date{June 2025}

\begin{document}

\maketitle
\begin{abstract}
Recent works~\cite{ben2023resistance,ratliff2024framework} have started to theoretically investigate how we can protect differentially private programs against timing attacks, by making the joint distribution the output and the runtime differentially private (JOT-DP). However, the existing approaches to JOT-DP have some limitations, particularly in the setting of unbounded DP (which protects the size of the dataset and applies to arbitrarily large datasets). First, the known conversion of pure DP programs to pure JOT-DP programs in the unbounded setting~\cite{ben2023resistance} (a) incurs a constant additive increase in error probability (and thus does not provide vanishing error as $n\to\infty$) (b) produces JOT-DP programs that fail to preserve the computational efficiency of the original pure DP program and (c) is analyzed in a toy computational model in which the runtime is defined to be the number of coin flips. For approximate JOT-DP, an efficient conversion with vanishing error in the RAM model is known~\cite{haeberlen2011differential,ratliff2024framework}, but only applies to programs that run in $O(n)$ time on datasets of size $n$, as linear runtime is implied by ``timing stability,'' the timing analogue of global sensitivity. In this work, we overcome these limitations. Specifically:
\begin{enumerate}
    \item We show that the error required for pure JOT-DP in the unbounded setting depends on the model of computation.
    \begin{itemize}
        \item In a randomized RAM model where the dataset size $n$ is given (or can be computed in constant time) and we can generate random numbers (not just random bits) in constant time, polynomially small error probability is necessary and sufficient. 
        \item If $n$ is not given or we only have a random-bit generator, an (arbitrarily small) constant error probability is necessary and sufficient. 
    \end{itemize}
    \item The aforementioned positive results are proven by efficient procedures to convert any pure JOT-DP program $P$ in the upper-bounded setting to a pure JOT-DP program $P'$ in the unbounded setting, such that the output distribution of $P'$ is $\gamma$-close in total variation distance to that of $P$, where $\gamma$ is either an arbitrarily small constant or polynomially small, depending on the model of computation. 
\end{enumerate}

\end{abstract}
\newpage
\tableofcontents
\newpage
\input{intro}

\input{prelims}

\input{pure-timing-private-construction}

\input{lower-bound}
\input{ram-timing-private}

\bibliography{refs}
\appendix 
\input{appendix}
\end{document}

%% file: intro.tex
\section{Introduction}
\emph{Timing side-channel attacks}, in which an attacker measures the runtime of an algorithm to infer otherwise private information, have proven to be a highly effective method for breaking modern cryptographic systems. These attacks have been used to extract signing keys from cryptographic processors~\cite{kocher1996timing}, recover private keys from remote TLS servers~\cite{brumley2005remote,brumley2011remote,al2013lucky,albrecht2016lucky}, and leak sensitive information from hidden hardware states (e.g., Spectre~\cite{kocher2020spectre} and Meltdown~\cite{lipp2018meltdown}).

Early research on timing attacks focused on identifying and mitigating timing side-channels in cryptographic implementations. However, the growing adoption of \emph{differential privacy} (DP)~\cite{dwork2006calibrating}—an influential framework for privacy-preserving data analysis, used in applications such as analyzing user behavior~\cite{messing2020facebook,greenberg2016apple} and releasing aggregate statistics~\cite{abowd2018us,aktay2020google}—has introduced new avenues for timing attacks. Differential privacy achieves its protections by adding carefully calibrated noise to computations, ensuring that the output of a query is not too sensitive to the addition or removal of a single data point. However, researchers have already identified that the runtime of user-defined queries~\cite{haeberlen2011differential,andrysco2015subnormal} and noise samplers~\cite{jin2021we} can severely violate the promises of differential privacy. Thus, recent research has begun to theoretically investigate \emph{timing-private} DP programs, which ensure that the joint distribution of the output and runtime satisfies differential privacy~\cite{ben2023resistance,ratliff2024framework}.

\begin{definition}[$(\varepsilon, \delta)$-Joint Output/Timing Privacy~\cite{ben2023resistance,ratliff2024framework}]
    \label{def:approx-joint-privacy}
         Let $\mathcal{X}$ be a dataset space with an adjacency relation, $\mathcal{E}$ be a set of execution environments for a computational model, $\mathcal{Y}$ be an output space, $\mathcal{T} \subseteq \mathbb{R}^{\ge 0}$ a set of possible runtimes in the model, and $P: \mathcal{X}\times\mathcal{E} \rightarrow \mathcal{Y}\times\mathcal{E}$ a randomized program in the model. Then we say that $P$ is $(\varepsilon, \delta)$-\emph{jointly output/timing-private} (JOT-DP) if for all adjacent $x, x' \in \mathcal{X}$, all pairs of input-compatible execution environments $\env, \env' \in \mathcal{E}$, and all $S \subseteq \mathcal{Y}\times\mathcal{T}$
    \begin{flalign*}
        \Pr[(Y, T) \in S] \le e^\varepsilon\cdot  \Pr[(Y', T') \in S] + \delta
    \end{flalign*}

    where $Y = \out(P(x, \env))$ and $\TP{x, \env}$ denote the output and runtime of $P$ respectively (and defined similarly for $Y'$ and $T'$). When $\delta = 0$, we say that $P$ achieves \emph{pure} JOT-DP, and when $\delta > 0$, we say that $P$ achieves \emph{approximate} JOT-DP. The dataset space $\mathcal{X}$ and its adjacency relation determine whether the program achieves JOT-DP in the \emph{bounded} (the number of records $n = |x|$ in a dataset $x \in \mathcal{X}$ is fixed and publicly known), \emph{unbounded} ($|x|$ can be arbitrarily large and is intended to remain private), or\emph{upper-bounded} (upper bound on $|x|$ is publicly known, but $|x|$ should remain private subject to the upper bound)  settings. 
    \end{definition}

The above notion of timing privacy requires that the joint distribution of the program $P$'s output and runtime when executed in the environment $\env$ satisfies the standard definition of differential privacy. Informally, the execution environment represents the state of the computer before execution, for example, the contents of memory. We give further background on execution environments in Section~\ref{section:preliminaries}. 

Beyond its role in data analysis, DP has recently emerged as a valuable tool for designing more practical and efficient cryptographic systems. For instance, metadata-private messaging systems have leveraged weaker DP guarantees to improve performance over purely cryptographic approaches~\cite{lazar2018karaoke,van2015vuvuzela}. Similarly, the concept of \emph{differential obliviousness}, which extends DP-like guarantees to the memory access patterns of a process, has been proposed as a lightweight alternative to oblivious RAM~\cite{chan2022foundations}. Apple has even adopted DP in their implementation of \emph{enhanced visual search} as a more efficient method of anonymizing user queries~\cite{apple2024homomorphic,asi2024scalable}. Thus, there is a pressing need to understand the extent to which DP implementations can be protected from timing attacks.

\subsection{Limitations of Prior Approaches}
One general technique suggested by Haeberlen, Pierce, and Narayan~\cite{haeberlen2011differential} and formalized and generalized by Ratliff and Vadhan~\cite{ratliff2024framework}, for constructing JOT-DP programs involve first establishing a bound on the program's \emph{timing stability}, which is the timing equivalent of global sensitivity. Once this bound is established, the output release is delayed by an amount sampled from a distribution, scaled according to the timing stability and the desired privacy level. However, this technique seems to inherently result in \emph{approximate} privacy $(\delta > 0)$ because it can only add non-negative ``noise'' to the program's runtime. It is possible to achieve \emph{pure} $(\delta = 0)$ JOT-DP in the bounded-DP (or upper-bounded DP) setting, where the maximum size of the input is known and public, by padding execution to the worst case runtime. However, in the \emph{unbounded} DP setting, where the size of the program's input is unknown and must be protected by differential privacy, achieving JOT-DP for non-trivial programs introduces several interesting challenges.

First, Ben Dov, David, Naor, and Tzalik~\cite{ben2023resistance} show that pure DP programs in the unbounded setting can be converted into JOT-DP programs, but the conversion results in a constant additive increase in error probability (and thus does not provide vanishing error as the dataset size $n\to\infty$). It remains unknown whether one must tolerate this constant probability of constant additive error, or if it is possible to construct pure JOT-DP programs that maintain the same asymptotic accuracy guarantees as their non-timing-private versions (which typically have error probability that tends to $0$ as the dataset size $n$ grows). Furthermore, the conversion technique of~\cite{ben2023resistance} can produce in a JOT-DP program that is inefficient, even if the original pure DP program is efficient. The reason being that constructing the sampler for the JOT-DP program requires computing the probabilities of all possible outputs of the original program, which may take exponential time even if the original program runs in expected polynomial time. 

For approximate JOT-DP, the previously mentioned technique of adding random delays can achieve vanishing error, but it is only applicable to programs that are timing stable, which implies an $O(n)$ runtime on datasets of size $n$. Consequently, in the unbounded DP setting, this approach cannot ensure timing privacy for common DP algorithms with superlinear runtimes. For example, the \emph{smooth-sensitivity median} algorithm~\cite{nissim2007smooth} for differentially private median computation requires sorting the input data, which incurs a $\Omega(n \log n)$ runtime for any comparison-based sorting algorithm. Thus, in the unbounded DP setting, no method is currently known for constructing even \emph{approximately} JOT-DP analogues of DP algorithms that require superlinear runtime.

\subsection{Our Results}
In this work, we address the above limitations and introduce new techniques for constructing differentially private programs in the unbounded setting that are secure against timing attacks. We adopt the RAM model of computation with a random number generator, which we view as more natural than the model used by Ben Dov et al.~\cite{ben2023resistance} (see Section~\ref{section:preliminaries}). Their results used a toy computational model equipped solely with a random bit generator, where the total runtime was measured by the number of random bits generated before halting. Additionally, since we focus on the unbounded setting of differential privacy, we do not utilize the Word RAM model, as its finite word size\footnote{The Word RAM model sometimes defines the word size $\omega$ as $\omega = O(\log n)$, where $n$ is the program's input length. However, this introduces additional complexities, as the word size itself may reveal information about the input length.} imposes constraints on addressable memory and, consequently, on input size. However, we note that our transformations do not rely on the RAM model’s capability to generate integers that are super-polynomially large in the program's runtime.

Within the RAM model of computation, we answer an open question by Ben Dov et al.~\cite{ben2023resistance} and construct an efficient procedure for converting any pure DP program $P$ in the \emph{upper-bounded} setting (see Definition~\ref{def:approx-joint-privacy}) to a pure JOT-DP program $P'$ for the unbounded setting whose output distribution differs by at most a constant $\beta$ in total variation distance from $P$. 

\begin{reptheorem}{theorem:constructive}[Pure JOT-DP RAM programs in the Unbounded Setting]
      For all $\beta > 0$, $\varepsilon > 0$, $\varepsilon' > \varepsilon$, and $\varepsilon$-JOT-DP RAM programs $P:\mathcal{X}\times\mathcal{E}\to\mathcal{Y}\times\mathcal{E}$ in the \emph{upper}-bounded setting, there exists a $\varepsilon'$-JOT-DP RAM program $P':\mathcal{X}\times\mathcal{E}\to\mathbb{N}\times\mathcal{E}$ for the unbounded setting such that
    \begin{align*}
        \Big\|\out(P(x, \env)) - \out(P'(x,\env)) \Big\|_{TV} < \beta
    \end{align*}
   Furthermore, the mechanism $P'(x, \env)$ is simply an explicit algorithm that makes one oracle call to $P$ on a truncation of $x$ along with an additional computation that takes time $O(|x|)$ with high probability.
\end{reptheorem}

Given the above result, we can convert pure JOT-DP programs with superlinear runtime in the upper-bounded setting into pure JOT-DP programs in the unbounded setting.  Note that the requirement that $P$ has a pure JOT-DP implementation in the upper-bounded setting is not overly restrictive, as many useful programs have such implementations (for example, by padding all executions to the maximum runtime on datasets of size within the upper bound). This resolves the open problem of Ben Dov et al.~\cite{ben2023resistance}. 

Note that Theorem~\ref{theorem:constructive} allows for a constant additive error in the output with constant probability, even as $n \to \infty$. This raises the question of whether such an error is unavoidable, a question we address in this work. Specifically, we show that the answer depends on the computational model starting with the following lower bound:

\begin{reptheorem}
{theorem:lower-bound}[Lower Bound for Pure JOT-DP Programs]
    Let $P:\mathcal{X}\times\mathcal{E}\to\mathbb{N}\times\mathcal{E}$ be a $\varepsilon$-JOT-DP program in the unbounded setting for some computational model, and that releases an estimate of its true input length. Let $t_0 = O(1)$ be the expected runtime of $P$ on the empty dataset, and $p_{t_0}(n) \le 1$ be the smallest non-zero probability such that $Z\sim Bernoulli(p_{t_0}(n))$ can be sampled in time $2\cdot t_0$ when $P$ is executed on a dataset consisting of $n$ records equal to a fixed value (e.g., $0$). Then $\exists c$ such that $\forall x\in\mathcal{X}, |x| \ge c$:
    \begin{align*}
     \Pr\Big[\Big|\out(P(x, \env)) - n\Big| > n - c\Big] > p_{t_0}(n)
    \end{align*}
    for $n = |x|$ as $n\to\infty$.
\end{reptheorem}

The above theorem gives a \emph{computational-model-dependent} lower bound on the achievable utility for pure JOT-DP programs. In particular, with probability at least $p_{t_0}(n)$, the program will output a useless result (since the error is of the order $O(n)$). If our computational model includes only a coin-tossing function as in the BDNNT model (Section~\ref{section:preliminaries}), then $p(n) \ge 2^{-2\cdot t_0}$, so the error probability is at least a constant. On the other hand, we show that in a randomized RAM model of computation where the input length is given as part of the input, we can achieve error that decays inverse polynomially in the input length (Section~\ref{sec:lower-bound}).

\begin{reptheorem}{theorem:dp-sum-unbounded}[Pure JOT-DP RAM Programs in the Unbounded Setting]
    For all $\varepsilon > 0$, $\varepsilon' > \varepsilon$, and $\varepsilon$-DP RAM programs $P:\mathcal{X}\times\mathcal{E}\to\mathcal{Y}\times\mathcal{E}$ in the upper-bounded setting, there exists a $\varepsilon'$-JOT-DP RAM program $P':\mathcal{X}\times\mathcal{E}\to\mathbb{N}\times\mathcal{E}$ for the unbounded setting satisfying $\forall x \in\mathcal{X}$ of length $n$, $\forall$ input-compatible $\env\in\mathcal{E}$, $\forall c \ge 2$:
    \begin{align*}
        \Big\|\out(P(x, \env)) - \out(P'(x, \env))\Big\|_{TV} < O\left(\frac{1}{n^c}\right)
    \end{align*}
    Furthermore, the mechanism $P'(x, \env)$ is an explicit algorithm that makes one oracle call to $P$ on a truncation of $x$ along with an additional computation that takes time $O(|x|)$ with high probability.
\end{reptheorem}

In particular, we obtain the following pure JOT-DP implementation of the Laplace mechanism. 

\begin{repcorollary}{corollary:jot-dp-laplace}[Pure JOT-DP Laplace Mechanism]
    For all $\varepsilon > 0$, and all $c > 0$, there exists an $\varepsilon$-JOT-DP RAM program $P:\mathcal{X} \times \mathcal{E} \to \mathcal{Y} \times \mathcal{E}$ for releasing sums in the unbounded setting such that for all datasets $x \in \mathcal{X}$:
    \begin{align*}
        \Pr\left[\Big|\out(P(x, \env)) - \sum x_i\Big| \geq \frac{C\cdot \ln(n)}{\varepsilon}\right] < O\left(\frac{1}{n^c}\right)
    \end{align*}
    where $n = |x|$ is the size of the dataset, and $C > 0$ is a universal constant.
\end{repcorollary}

Theorem~\ref{theorem:dp-sum-unbounded} shows that we can construct \emph{pure} JOT-DP sums in the randomized RAM model that achieve comparable privacy guarantees to the standard $\varepsilon$-DP Laplace mechanism. While the error probability is not exponentially decaying as in the standard (non-timing-private) Laplace mechanism, the JOT-DP mechanism still ensures that the error vanishes at a polynomial rate. This is an improvement over the constant error bound given by Ben Dov et al.~\cite{ben2023resistance} and Theorem~\ref{theorem:constructive}, and is the best possible by Theorem~\ref{theorem:lower-bound} since $p_{t_0}(n) \ge n^{-2^{2\cdot t_0}}$ (Section~\ref{sec:lower-bound}). 

\subsection{Techniques}
We introduce a new technique for constructing \emph{pure} JOT-DP programs in the unbounded setting (Section~\ref{sec:pure-timing-privacy}). Specifically, both Theorem~\ref{theorem:constructive} and Theorem~\ref{theorem:dp-sum-unbounded} are obtained by (a) constructing a pure JOT-DP program that coarsely estimates its input length, (b) truncating the dataset based on this estimate, and (c) running an upper-bounded pure JOT-DP program using the estimate as the upper bound on dataset size.

Informally, the coarse estimation procedure repeatedly queries whether the input length exceeds some threshold, where the threshold doubles in each round, continuing until the answer is ``no.'' Because the query responses must satisfy differential privacy, each has an associated error probability. However, this error can be made arbitrarily small.

To achieve the bound for RAM programs in Theorem~\ref{theorem:dp-sum-unbounded}, we leverage the fact that the input length $n$ is known in advance, allowing the error parameter to explicitly depend on $n$. Specifically, the algorithm uses an adaptive sampling procedure that repeatedly flips a biased coin, starting with an initial success probability of $1/(n + k)^c$ for constants $c$ and $k$. After each unsuccessful flip, the success probability is updated to $1/(n - i + k)^c$ until it reaches a fixed probability of $1/k^c$. Flipping continues until the first success, at which point the algorithm outputs the total number of flips. Importantly, this ensures (a) the runtime is exactly determined by the output, and (b) the output itself is an $\varepsilon$-DP estimate of the input length, making the algorithm $\varepsilon$-JOT-DP. Using this adaptive sampler as a building block, we construct pure JOT-DP RAM programs that achieve the bound from Theorem~\ref{theorem:dp-sum-unbounded}.  Our construction is optimal in terms of its failure probability (i.e., it achieves error probability that vanishes at a polynomial rate). In particular, Theorem~\ref{theorem:lower-bound} establishes that this decay rate is the best achievable within the RAM model, and in fact, weaker computational models can only achieve a constant error probability. 

To establish the lower bound in Theorem~\ref{theorem:lower-bound}, we analyze the output support of a pure DP program in the unbounded setting conditioned on halting within time $t \le 2t_0$, where $t_0$ is the expected runtime of the program on the empty input. Our argument relies on the observation that even when the database is very long, the mechanism must with some probability behave exactly as it does on the empty database. On the empty database the mechanism halts within time $2\cdot t_0$ with high probability, and thus the same behavior must occur with positive probability on the long database. By the definition of JOT-DP, the output support conditioned on halting within this time bound must be finite and identical across all inputs. We further show that every output in this finite support provides poor utility for most inputs. Nonetheless, the mechanism must output an element from this finite set with probability at least $p_{t_0}(n)$, where $p_{t_0}(n)$ is the smallest non-zero probability computable by a program running in time $2t_0$. Together, our results demonstrate that in many reasonable computational models, pure JOT-DP programs in the unbounded setting necessarily incur an added utility cost to maintain their timing privacy guarantees.

%% file: prelims.tex
\section{Preliminaries}\label{section:preliminaries}

\textbf{RAM Model}. Throughout this work, we will use the idealized RAM model of computation. The RAM model consists of an infinite sequence of memory cells, each capable of storing arbitrarily large natural numbers. Variables are stored in registers and RAM programs can perform a set of basic operations for arithmetic (addition, subtraction, multiplication, and integer division), Boolean logic (AND, OR, NOT), and reading/writing memory.  They also allow conditional jumps (e.g., \textbf{if} CONDITIONAL \textbf{goto} LINE), which implement standard control flow constructs such as \textbf{if} and \textbf{if/else} statements. Additionally, we allow our RAM programs to use randomness by executing a $\texttt{RAND}(n)$ instruction to uniformly sample an integer from $\{0, \dots, n\}$. The runtime of a RAM program is defined as the number of basic instructions\footnote{We count conditional branching instructions and randomness sampling as basic instructions taking one time step.} executed before the program halts, meaning that the set of possible runtime values $\mathcal{T}$ corresponds to the natural numbers $\mathbb{N}$. We also include a built-in variable $\texttt{input\_len}$ and $\texttt{input\_ptr}$ (respectively $\texttt{output\_len}$ and $\texttt{output\_ptr})$, which stores the length of the program's input and its location in memory respectively (and similarly for the program's output). 

We also consider a variant of the randomized RAM model where the program lacks direct access to the input length (i.e., \texttt{input\_len} is not initialized). This model is strictly weaker than the one described above, as the program must compute the input length during execution if needed. In this model, the end of the input is marked in memory by a special delimiter symbol reserved for indicating the end of the input. \\

\noindent\textbf{RAM$_{\textrm{BDDNT}}$}. Earlier work by Ben Dov et al.\cite{ben2023resistance} used a computational model where randomness is provided by a simple coin-tossing function (i.e., a draw from Bernoulli($1/2$)), and runtime is measured solely by the number of coin tosses performed. This is in contrast to our RAM model, where a call to \texttt{RAND}(n) samples a random integer from $\{0,\dots, n\}$ and runtime is measured by the total number of instructions executed before halting. Consequently, when discussing the main results of Ben Dov et al. in Section~\ref{sec:pure-timing-privacy}, we denote by RAM$_{\textrm{BDDNT}}$ the RAM model in which randomness is restricted to calls to a simple Bernoulli($1/2$) sampler and runtime is measured by the total number of such calls.\\

\noindent\textbf{Execution Environments.} The runtime of a program on a given input is often highly dependent on the program's \emph{execution environment}. For example, a program executing on x86 hardware will have runtime that can be influenced by concurrent processes executing on the system, the state of the branch predictor (if the hardware supports speculative execution), the cache state, and various other forms of resource contention that might occur on the system. Thus, we always consider a program's runtime to be a function of both the input and some execution environment $\env \in \mathcal{E}$ representing the machines initial state before executing the program. For example, the set of execution environments $\mathcal{E}$ for RAM programs includes all possible memory configurations and initial values of built-in variables. 

\begin{definition}[RAM Execution Environment]
The \emph{execution environment} $\env$ of a RAM program is the infinite sequence $(v_{0}, v_{1}, \dots, )$ such that $M[i] = v_i$ for all $i$ along with the values stored by the built-in variables such as $\texttt{input\_ptr}$ and $\texttt{input\_len}, \texttt{output\_ptr},$ and $\texttt{output\_len}$. 
\end{definition}

We note that while the execution environment significantly influences program runtimes in practice, throughout this paper, we will work with RAM programs whose runtime and output distributions are jointly independent of the execution environment. Additionally, we emphasize that not every execution environment is \emph{compatible} with a given input. For example, a RAM program that processes a length-$n$ string consisting entirely of $1$'s may be incompatible with a RAM environment where memory has been zeroed out. We say that $P(x, \env)$ is undefined for such environments $\env$ that are incompatible with an input $x$. Consequently, our definitions are often quantified over all pairs of inputs $x \in \mathcal{X}$ and \emph{input-compatible} execution environments $\env \in \mathcal{E}$. \\


\noindent\textbf{The Bounded, Upper-Bounded, and Unbounded Settings.} 
Differential privacy is defined with respect to an input space $\mathcal{X}$, where inputs typically consist of $n \geq 0$ records drawn from some domain $\mathcal{D}$. Formally, we define the input space as $\mathcal{X} = \bigcup_{n=0}^\infty \mathcal{D}^n$. Two inputs $x, x' \in \mathcal{X}$ are said to be \emph{adjacent} with respect to a dataset distance metric $d_{\mathcal{X}}$ if $d_{\mathcal{X}}(x, x') \leq 1$. 

We can distinguish between different settings of differential privacy according to whether the program's input size is considered public information. In the \emph{bounded} setting, the input length is assumed to be publicly known. Consequently, privacy guarantees do not extend to hiding the dataset size. This assumption simplifies the defense against timing attacks, as execution can often be padded to match the worst-case runtime for inputs of a given length. A stronger flavor of DP is the \emph{upper-bounded} setting, where only an upper bound $n_{\rm{max}}$ on the program's input length is assumed to be public. This setting is commonly encountered in practice, as one can typically establish a conservative upper limit on the dataset size. Similarly to the bounded setting, one can often pad execution up to a worst case runtime\footnote{However, in practice, this approach may be inefficient, as the program's worst-case runtime for inputs of length $n_{\rm{max}}$ could be prohibitively long.} in the upper-bounded setting to protect against timing attacks. However, the strongest privacy guarantee is provided by the \emph{unbounded} setting, in which the input length is intended to be kept private and can be arbitrarily large. 

In both the upper-bounded and unbounded settings, adjacency is often defined using an \emph{insert-delete} distance.

\begin{definition}[Insert-Delete Distance]
For $x \in \mathcal{D}^*$, an insertion to $x$ is an addition of an
element $z$ to a location in $x$ resulting in a new input $x' = [x_1, \dots, x_i, z, x_{i+1}, \dots, x_n]$.
Likewise, a deletion from $x$ is the removal of an element from a location $i$, giving a new input $x' = [x_1, \dots, x_{i - 1}, x_{i + 1}, \dots, x_n]$. We define the \emph{insert-delete distance}, denoted $\did$, of inputs $x, x' \in \mathcal{D}^*$ to be the minimum number of insertion and deletion operations needed to transform $x$ into $x'$.
\end{definition}

Throughout this paper, we operate within the unbounded setting of differential privacy and will therefore implicitly use $\did$ as the adjacency relation in all definitions and theorems. \\

\noindent\textbf{Properties of JOT-DP Programs.} We will frequently use the fact that DP programs with constant-time execution on all inputs are also JOT-DP. 

\begin{lemma}[Constant-Time JOT-DP Programs~\cite{ratliff2024framework}]\label{lemma:constant-time-programs}
     If a program $P:\mathcal{X}\times\mathcal{E}\to\mathcal{Y}\times\mathcal{E}$ is $\varepsilon$-DP in its output and there exists a constant $c$ such that $\TP{x, \env} = c$ for all $x\in\mathcal{X}$ and $\env\in\mathcal{E}$, then $P$ is $\varepsilon$-JOT-DP.
\end{lemma}

We will also make use of programs $P$ that are the composition of JOT-DP programs $P_1$ and $P_2$. Such programs $P$ are constructed by \emph{chaining} together $P_1$ such that its output along with its unaltered input are fed as the input to program $P_2$. 

\begin{lemma}[Sequential Composition of JOT-DP RAM Programs~\cite{ratliff2024framework}]\label{lemma:composition-timing-privacy}
Let $P_1 : \mathcal{X}\times\mathcal{E}\to\mathcal{Y}\times\mathcal{E}$ be an $\varepsilon_1$-JOT-DP RAM program. Let $P_2: \mathcal{X}\times\mathcal{E}\to\mathcal{Z}\times\mathcal{E}$ be an $\varepsilon_2$-JOT-DP RAM program. Then the sequentially composed program $P_2 \otimes P_1: \mathcal{X}\times\mathcal{E}\to(\mathcal{Y}\times\mathcal{Z})\times\mathcal{E}$ that executes $P_1$ on the input $x$ followed by $P_2$ on input $x$ is $(\varepsilon_1 + \varepsilon_2)$-JOT-DP.
\end{lemma}


Finally, we review the Discrete Laplace distribution. 

\begin{definition}[Discrete Laplace Distribution]\label{def:discrete-laplace}
The \emph{Discrete Laplace distribution} with shift $\mu \in \mathbb{N}$ and scale $s > 0$, is supplied a $x \in \mathbb{Z}$, has the probability mass function:
\[
p(x \mid \mu, s) = \frac{e^{1/s} - 1}{e^{1/s} + 1} \cdot e^{-|x - \mu|/s}
\]
The cumulative distribution function (CDF) is given by:
\[
F(x \mid \mu, s) = 
\begin{cases} 
\frac{e^{1/s}}{e^{1/s} + 1} \cdot e^{-(\mu - x)/s}, & \text{if } x \leq \mu, \\
1 - \frac{1}{e^{1/s} + 1} \cdot e^{-(x - \mu)/s}, & \text{if } x > \mu.
\end{cases}
\]
\end{definition}

Throughout this paper, we will often use a \emph{censored}\footnote{The literature also refers to this as the Truncated Geometric Mechanism~\cite{ghosh2012universally}.} version of the Discrete Laplace distribution, which we denote by $\texttt{CensoredDiscreteLaplace}(\mu, s, \ell, u)$. This distribution is additionally parameterized by lower and upper bounds $\ell$ and $u$ so that $X \sim \textrm{DiscreteLaplace}(\mu, s)$ is clamped to the range $[\ell, u]$. 

\begin{lemma}[Censored Discrete Laplace is DP~\cite{ghosh2012universally}]
\label{lemma:censored-laplace}
    Let $x, x'$ be adjacent datasets, $\ell, u \in \mathbb{N}$ for $\ell \le u$, $f: \mathcal{X}\to\mathbb{Z}$ a function with global sensitivity $\Delta$, and $\varepsilon > 0$. Then $M(x) = \texttt{CensoredDiscreteLaplace}(\mu = f(x), s = \Delta/\varepsilon, \ell, u)$ is $\varepsilon$-DP.
\end{lemma}

We remark that constant-time instantiations of the censored Discrete Laplace distribution are known~\cite{balcer2019differential,ratliff2024framework}. Using Lemma~\ref{lemma:censored-laplace} and the fact that we can implement the censored Discrete Laplace mechanism to run in fixed time that depends only on $s$, $\ell$, and $u$, we obtain a JOT-DP version of $\texttt{CensoredDiscreteLaplace}$ (by Lemma~\ref{lemma:constant-time-programs}). 

\begin{lemma}[Censored Discrete Laplace is JOT-DP]
\label{lemma:censored-laplace-jot-dp}
    Let $x, x'$ be adjacent datasets, $\ell, u \in \mathbb{N}$ for $\ell \le u$, $f:\mathcal{X}\to\mathbb{Z}$ a function with global sensitivity $\Delta$, and $\varepsilon > 0$. Then there exists a RAM program $P: \mathcal{X}\times\mathcal{E}\to\mathcal{Y}\times\mathcal{E}$ such that $P(x,\env)$ samples from $\texttt{CensoredDiscreteLaplace}(\mu = f(x), s = \Delta/\varepsilon, \ell, u)$ in time $O(u)$ and achieves $\varepsilon$-JOT-DP.
\end{lemma}

%% file: pure-timing-private-construction.tex
\section{Pure Timing-Private Programs}\label{sec:pure-timing-privacy}
In this section, we characterize pure JOT-DP programs in the unbounded setting. To begin, we reintroduce a result from Ben Dov et al.~\cite{ben2023resistance} for the RAM$_{\textrm{BDDNT}}$ model of computation (described in Section~\ref{section:preliminaries}). 

\begin{algorithm}[t]
\begin{flushleft}

\textbf{Input:} A dataset $x$, privacy parameter $\varepsilon' > 0$, failure parameter $\beta > 0$, and $\varepsilon$-JOT-DP program $P$ for the upper-bounded setting.

\vspace{5px}
\textbf{Output:} The output of $P$ when executed on the input $x$ truncated to $\hat{n}$ records where $\hat{n}$ is a $\varepsilon'$-DP count on the size of $x$.

\end{flushleft}
\begin{algorithmic}[1]
\STATE $\varepsilon' = \varepsilon' / 2$;
\STATE $\beta' = \beta / 2$;
\STATE $m = \frac{2}{\varepsilon'}\cdot\lceil\ln (\frac{1}{\beta'})\rceil$;
\WHILE{$\textbf{True}$}
    \STATE $\textrm{Scan first } m \textrm{ rows of input and set } \mu = \min\{n, m\}$; \COMMENT{note that $n = |x|$}
    \STATE $s = 1 / \varepsilon'$;
    \STATE $\ell = 0$; \COMMENT{lower bound for censored Discrete Laplace mechanism}
    \STATE $u = m$; \COMMENT{upper bound for censored Discrete Laplace mechanism}
    \STATE $\hat{n} = \texttt{CensoredDiscreteLaplace}(\mu, s, \ell, u)$; 
    \IF{$\hat{n}< \frac{m}{2}$}
        \STATE $s = 1/\varepsilon$;
        \STATE $\hat{x} = \texttt{Truncate}(x, m)$; \COMMENT{Truncates the dataset to $m$ records if $|x| > m$ }
        \STATE $\textrm{run } P^m(\hat{x})$; \COMMENT{$P$ in the $m$-upperbounded setting}
        \RETURN $\texttt{output of } P(\hat{x})$;
    \ELSE
        \STATE $\varepsilon' = \varepsilon' / 2$;
        \STATE $\beta' = \beta' / 2$;
        \STATE $m = \frac{2}{\varepsilon'}\cdot \lceil\ln (\frac{1}{\beta'})\rceil$;
    \ENDIF
\ENDWHILE
\end{algorithmic}

\caption{General Construction for Pure JOT-DP Programs in the Unbounded Setting}\label{program:cdl-mechanism}
\end{algorithm}

\begin{theorem}[Ben Dov et al.~\cite{ben2023resistance}]\label{theorem:ben-dov-main}
    Let $P : \mathcal{X}\times\mathcal{E}\to\mathcal{Y}\times\mathcal{E}$ be any $\varepsilon$-DP RAM$_{\textrm{BDDNT}}$ program. For any $\beta > 0$, $\varepsilon' > \varepsilon$ there exists a $\varepsilon'$-JOT-DP RAM$_{\textrm{BDDNT}}$ program $P': \mathcal{X}\times\mathcal{E}\to\mathcal{Y}\times\mathcal{E}$ such that for all $x \in \mathcal{X}$, $\env \in \mathcal{E}$:
    \begin{align*}
        \Big\|\out(P(x, \env)) - \out(P'(x, \env))\Big\|_{TV} < \beta
    \end{align*}
\end{theorem}

Theorem~\ref{theorem:ben-dov-main} is promising in that it suggests one can achieve pure JOT-DP as long as one is willing to tolerate a constant error probability. However, the authors left as an open question whether the pure JOT-DP program $P'$ can maintain the computational efficiency of $P$. We give a positive answer to this question for a wide class of programs (namely, programs that have pure JOT-DP implementations in the upper-bounded setting).

We start with the construction described by Program~\ref{program:cdl-mechanism}, which first computes a pure JOT-DP \emph{over}-estimate of a program's input length that holds with arbitrarily small probability $\beta$. Intuitively, Program~\ref{program:cdl-mechanism} scans its input $m_i$ entries at a time, where $m_i$ depends only on the parameters $\varepsilon'$, $\beta$, and $i$, and $m_{i+1}$ is roughly $2m_i$. In each iteration, the program releases a DP count on the first $m_i$ entries of the input (i.e., a DP count on $m_i$ if $m_i < |x|$, and a DP count on $|x|$ otherwise) and stops scanning the input once it releases a count less than $m_i/2$. This step consumes $\varepsilon'$ of the privacy budget to generate the coarse estimate $m_k$, which serves as an upper bound on the input size with high probability. Once this estimate is determined, the input is truncated (if necessary) in time that depends only on the differentially private overestimate $m_k$. Finally, the program $P^m$ is executed on the transformed dataset $\hat{x}$, where $P^m$ is the program $P$ restricted to inputs of length at most $m_k$.

\begin{lemma}
\label{lemma:cdl-mechanism-privacy}
Let $P':\mathcal{X}\times\mathcal{E}\to\mathbb{N}\times\mathcal{E}$ be the RAM program implementing Program~\ref{program:cdl-mechanism} for a RAM program $P:\mathcal{X}\times\mathcal{E}\to\mathcal{Y}\times\mathcal{E}$ that is $\varepsilon$-JOT-DP in the upper-bounded setting. Then $P'$ achieves $(\varepsilon + \varepsilon')$-JOT-DP in the unbounded setting.
\end{lemma}
\begin{proof}
We can interpret the program $P'$ as the composition of two subprograms: $P_{\texttt{(line 13)}}$, which encompasses the execution of $P'$ up to line 13, and $P^m$, which represents the execution of $P$ on $\hat{x}$ within the upper-bounded setting, where inputs are guaranteed to have length at most $m$. By assumption, $P^m$ satisfies $\varepsilon$-JOT-DP. Furthermore, $\texttt{Truncate}$ is $1$-stable\footnote{A function $f:\mathcal{X}\to\mathcal{X}$ that maps datasets to datasets is $1$-stable if for all $x, x' \in\mathcal{X}$, $\did(f(x), f(x')) \le c\cdot\did(x, x')$.}, and therefore for all $x, x'\in\mathcal{X}$ satisfying $\did(x, x') \le 1$, it follows that $\did(\texttt{Truncate}(x, m_k), \texttt{Truncate}(x', m_k)) \le 1$. Thus, to establish the overall privacy guarantee, it remains to show that $m$ is computed in a differentially private manner and that $P_{\texttt{(line 13)}}$ satisfies $\varepsilon'$-JOT-DP. Given these conditions, we can view the execution of $P'(x,\env)$ as the sequential composition of two JOT-DP programs $P_{\texttt{(line 13)}}(x, \env)$ and $P^m(\hat{x}, \hat{\env})$.

Observe that $P_{\texttt{(line 13)}}$ scans the input $m$ entries at a time, where $m$ depends only on $\varepsilon'$ and $\beta'$ (lines 3 and 18). At each iteration $i$ of the loop (for $i = 1\dots k$), let $\varepsilon_i = \varepsilon'$, $\beta_i = \beta'$, and $m_i = m$. During each loop iteration, the program invokes a censored Discrete Laplace mechanism to compute a $\varepsilon_i$-DP count over the first $m_i = 2\cdot \lceil \ln(1/\beta_i) \rceil / \varepsilon_i$ entries of the input where $\varepsilon_i = \varepsilon'/2^{i}$ and $\beta_i = \beta/2^{i}$. Recall that each invocation of the censored Discrete Laplace mechanism is $\varepsilon_i$-JOT-DP (Lemma~\ref{lemma:censored-laplace-jot-dp}). Thus, up to line $11$, the runtime of Program~\ref{program:cdl-mechanism} depends only on the number of iterations $k$, which is a post-processing function of the $k$ JOT-DP counts. Furthermore, once the condition on line 10 is met, the program performs a truncation operation, $\texttt{Truncate}(x, m)$. This operation executes in time dependent only on $m_k$ (e.g., by returning a copy of the first $m_k$ records of $x$).  Since $P'_{\texttt{(line 13)}}$ computes $k$ $\varepsilon_i$-DP counts (for $i = 1, \dots, k$) and runs in time determined solely by $k$, it follows that $P'_{\texttt{(line 13)}}$ satisfies $\varepsilon'$-JOT-DP since 

\begin{align*}
    \sum_{i=1}^{k} \varepsilon_i &\le \sum_{i=1}^{\infty} \varepsilon_i \\
    &= \sum_{i = 1}^\infty \frac{\varepsilon'}{2^{i}} \\
    &= \varepsilon' \cdot \sum_{i = 1}^\infty \frac{1}{2^{i}} \\
    &= \varepsilon' 
\end{align*}

Consequently, $P'$ is the composition of two JOT-DP programs, $P'_{\texttt{(line 13)}}$ and $P$. By the basic composition theorem (Lemma~\ref{lemma:composition-timing-privacy}), it follows that the overall program satisfies $(\varepsilon + \varepsilon')$-JOT-DP.
\end{proof}

\begin{lemma}
\label{lemma:cdl-mechanism-accuracy}
   Let $P':\mathcal{X}\times\mathcal{E}\to\mathbb{N}\times\mathcal{E}$ be the program implementing Program~\ref{program:cdl-mechanism} for a program $P:\mathcal{X}\times\mathcal{E}\to\mathcal{Y}\times\mathcal{E}$ that is $\varepsilon$-JOT-DP in the upper-bounded setting. Then $P'$ returns the output of $P$ on dataset $x$ with probability at least $1 - \beta$.
\end{lemma}
\begin{proof}
We first consider the case where Program~\ref{program:cdl-mechanism} terminates on loop iteration $k$ with $m_k < |x|$. In this scenario, $\hat{x} = \texttt{Truncate}(x, m_k)$ and $|\hat{x}| < |x|$ so the program returns the output of $P$ on the first $m_k$ entries of $x$. The probability of this occurring is given by

    \begin{align*}
        \Pr\left[\hat{n}_i
         < \frac{m_i}{2}\right] \le e^{- m_i\cdot\varepsilon_i / 2}\le \beta_i
    \end{align*}
    for $\hat{n}_i \sim \texttt{CensoredDiscreteLaplace}(\mu = m_i, s = 1/\varepsilon_i, \ell = 0, u = m_i)$ by the CDF of the Discrete Laplace distribution and the fact that $m_i \ge \frac{2}{\varepsilon_i}\cdot \ln(1/\beta_i)$. We union bound over the $k$ invocations of the mechanism:

    \begin{align*}
       \sum_{i = 1}^k \Pr\left[\hat{n}_i < \frac{m_i}{2}\right] &\le \sum_{i = 1}^\infty \beta_i \\
        &= \sum_{i = 1}^\infty \frac{\beta}{2^{i}} \\
        &= \beta \sum_{i = 1}^\infty \frac{1}{2^{i}} \\
        &= \beta
    \end{align*}

    We now condition on Program~\ref{program:cdl-mechanism} halting on loop iteration $k$ where $m_k \ge |x|$. In this case, $\hat{x} = x$ since $\texttt{Truncate}(x, m_k)$ does not alter $x$ when $m_k > |x|$. Thus $P'$ returns the output of $P$ executed on $x$ and the claim follows. 
\end{proof}

We now prove that for every pure JOT-DP program in the upper-bounded setting, we can obtain a pure JOT-DP program in the unbounded setting.

\begin{theorem}[Pure JOT-DP in the Unbounded Setting]
\label{theorem:constructive}
For all $0 < \beta < 1$, $\varepsilon > 0$, $\varepsilon' > \varepsilon$, and $\varepsilon$-JOT-DP RAM programs $P:\mathcal{X}\times\mathcal{E}\to\mathcal{Y}\times\mathcal{E}$ in the \emph{upper}-bounded setting, there exists a $\varepsilon'$-JOT-DP RAM program $P':\mathcal{X}\times\mathcal{E}\to\mathbb{N}\times\mathcal{E}$ for the unbounded setting such that
    \begin{align*}
        \Big\|\out(P(x, \env)) - \out(P'(x,\env)) \Big\|_{TV} < \beta
    \end{align*}
   Furthermore, the mechanism $P'(x, \env)$ is simply an explicit algorithm that makes one oracle call to $P$ on a truncation of $x$ along with an additional computation that takes time $O(|x|)$ with probability at least $1 - \beta - e^{-\Omega(|x|)}$.
\end{theorem}

\begin{proof}
    The proof follows from Lemma~\ref{lemma:cdl-mechanism-privacy} and Lemma~\ref{lemma:cdl-mechanism-accuracy}. In particular, given $P$, we construct the program $P'$ as described in Program~\ref{program:cdl-mechanism} so that $P'$ is $\varepsilon'$-JOT-DP for $\varepsilon' > \varepsilon$. What remains to be shown is that, with high probability, $P'$ runs in time $O(|x|)$ before making the oracle call to $P$.

    There are two cases. We first consider the case where $m_k < |x|$. As shown in Lemma~\ref{lemma:cdl-mechanism-accuracy}, this happens with probability: 

    \begin{align*}
        \sum_{i=1}^\infty \Pr\left[\hat{n}_i <  \frac{m_i}{2}\right] 
         &\le\sum_{i=1}^\infty  e^{- m_i\cdot\varepsilon_i / 2} \\
         &\le \sum_{i=1}^\infty \beta_i \\
         &= \beta
    \end{align*}

    where $\hat{n}_i \sim \texttt{CensoredDiscreteLaplace}(\mu = m_i, s = 1/\varepsilon_i, \ell = 0, u = m_i)$. Thus, we consider the case where $m_k > |x|$ and show that $P'$ stops looping when $m_k = O(|x|)$ with high probability. First, observe that
    \begin{align*}
        \frac{m_k}{m_{k-1}} &=  \frac{\frac{2^k}{\varepsilon'}\cdot \Big\lceil\log \frac{2^k}{\beta}\Big\rceil}{\frac{2^{k-1}}{\varepsilon'}\cdot \Big\lceil\log \frac{2^{k-1}}{\beta}\Big\rceil } \\
        &= 2\cdot \frac{r + 1}{r} \\
    \end{align*}

    for $r = k - 1 + \Big\lceil \log (1/\beta)\Big\rceil \ge 1$. Thus,

    \begin{align*}
        2 \le \frac{m_k}{m_{k-1}} \le 4
    \end{align*}

    Therefore, during each iteration of the loop, we increase the number of input entries scanned by at least a factor of $2$ and at most a factor of $4$. Now, consider the index $k$ such that $m_k > |x|$ but $m_{k-1} \le |x|$. Then $|x| < m_k \le 4\cdot |x|$. Furthermore, there exists a $j \ge k$ such that $4\cdot |x| \le m_j \le 16\cdot|x|$. Note that during this loop iteration we have $\hat{n}_j \sim \texttt{CensoredDiscreteLaplace}(\mu = |x|, s = 1/\varepsilon_j, \ell = 0, u = m_j)$ and 
    \begin{align*}
        \Pr\left[\hat{n}_j >  \frac{m_j}{2}\right] = 1 - F\left(\frac{m_j}{2} | \mu = |x|, s = 1/\varepsilon_j\right) \le e^{-\Omega(|x|)}
    \end{align*}

    where $F$ is the CDF of the Discrete Laplace distribution. Thus, with probability at least $1 - \beta - e^{-O(|x|)}$, $P'$ stops looping on iteration $j = O(\log |x|)$ where $m_j = O(|x|)$. The runtime of each loop iteration is dominated by the time it takes to sample from \texttt{CensoredDiscreteLaplace}$(\mu, s, \ell, u = m_i)$, which can be bounded by $O(m_i)$ (Lemma~\ref{lemma:censored-laplace-jot-dp}), where $u$ starts at $m_1$ and at least doubles during each iteration up to $m_j = O(|x|)$. It follows that, with high probability, $P'$ runs in time $O(|x|)$ before making the oracle call to $P$ and terminating.
\end{proof}

Theorem~\ref{theorem:constructive} shows that for any DP RAM program $P$, it is possible to efficiently construct a JOT-DP variant $P'$ with comparable privacy guarantees, incurring at most a $\beta$ increase in error probability. In other words, the modified program $P'(x)$ may return inaccurate results with probability at most $\beta$ higher than $P(x)$. However, the construction used in Theorem~\ref{theorem:constructive} does not extend to the $\text{RAM}_{\text{BDDNT}}$ model, as it relies on sampling from a Censored Discrete Laplace distribution using a constant number of coin flips. This sampling is possible only if all output probabilities are dyadic, but Lemma~\ref{lemma:no-dyadic-dl} shows that no nontrivial Censored Discrete Laplace distribution has this property. Consequently, even in the bounded DP setting, one cannot implement the standard Laplace mechanism with \emph{perfect} timing privacy guarantees, as can be done in the RAM model. Fortunately, we demonstrate that replacing the sampling from the Censored Discrete Laplace distribution in line~9 of Program~\ref{program:cdl-mechanism} with sampling from a closely related distribution whose probability masses are entirely dyadic ensures that Theorem~\ref{theorem:constructive} also holds in the $\text{RAM}_{\text{BDDNT}}$ model (see Appendix). 

Whether the constant additive increase in error probability  $\beta$ is unavoidable even in the RAM model remains an open question. In the next section, we establish a lower bound on the achievable utility of pure JOT-DP programs in the unbounded setting.

%% file: lower-bound.tex
\section{A Lower Bound for JOT-DP Programs}\label{sec:lower-bound}
A key question that arises from Theorem~\ref{theorem:constructive}  (Section~\ref{sec:pure-timing-privacy}) is whether the constant additive error in the output, which occurs with constant probability even as the input size $n$ grows, is unavoidable.  In this section, we address this question, exploring the inherent limitations of JOT-DP programs. We begin by establishing a lower bound, demonstrating that the achievable utility for pure JOT-DP programs depends on the computational model.

\begin{theorem}[Lower Bound for JOT-DP Counting Programs]\label{theorem:lower-bound}
    Let $P:\mathcal{X}\times\mathcal{E}\to\mathbb{N}\times\mathcal{E}$ be a $\varepsilon$-JOT-DP program in the unbounded setting for some computational model, and that releases an estimate of its true input length. Let $t_0 = O(1)$ be the expected runtime of $P$ on the empty dataset, and $p_{t_0}(n) \le 1$ be the smallest non-zero probability such that $Z\sim Bernoulli(p_{t_0}(n))$ can be sampled in time $2\cdot t_0$ when $P$ is executed on a dataset consisting of $n$ records equal to a fixed value (e.g., $0$). Then $\exists c$ such that $\forall x\in\mathcal{X}, |x| \ge c$:
    \begin{align*}
     \Pr\Big[\Big|\out(P(x, \env)) - n\Big| > n - c\Big] > p_{t_0}(n)
    \end{align*}
    for $n = |x|$ as $n\to\infty$.
\end{theorem}

\begin{proof}
    Let $P:\mathcal{X}\times\mathcal{E}\to\mathcal{Y}\times\mathcal{E}$ be a $\varepsilon$-JOT-DP program for releasing an estimate on the length of its input. Let $\mathbb{E}[\TP{\lambda, \env_\lambda}] = t_0$ where $t_0 = O(1)$ and $\lambda$ is the empty dataset and $\env_\lambda$ is an input-compatible execution environment. Let $p_{t_0}(n) \le 1$ be the smallest probability such that $Z\sim Bernoulli(p_{t_0}(n))$ is sampleable in time $2\cdot t_0$ when $P$ is executed on a dataset consisting of $n$ copies of a fixed record.

    Observe that $\Pr[\TP{\lambda, \env_\lambda} < 2\cdot t_0] \ge 1/2$ by Markov's inequality. Furthermore, let 
    \begin{align*}
        S = \supp(\out(P(\lambda, \env_\lambda))|\TP{\lambda, \env_\lambda} \le 2\cdot t_0)
    \end{align*}
    Note that $S$ is finite since $P(\lambda, \env_\lambda)$ can only generate a finite number of outputs within $2\cdot t_0$ time steps. We can set $c = \max_{y \in S} y$. Then $\forall n \ge c, y \in S$ it follows that 
    \begin{align*}
        |y - n| \ge n - c
    \end{align*}
    
    By pure JOT-DP, we have that for all $x$
    \begin{align*}
        \supp(\out(P(x, \env))|\TP{x,\env} \le 2\cdot t_0) = S
    \end{align*}
    Thus, we have $\Pr\left[|\out(P(x, \env)) - n| > n - c \right] > p_{t_0}(n)$ where $n = |x|$.
\end{proof}

Theorem~\ref{theorem:lower-bound} establishes that the achievable utility of pure JOT-DP programs in the unbounded setting depends on the computational model. For example, in the randomized RAM model, as we have defined it, where the program receives its input length $n$ as part of the input, then $p_{t_0}(n) \geq n^{-2^{2 \cdot t_0}}$ implying that $p_{t_0}(n)$ vanishes as $n \to \infty$. This can be realized, for instance, by a RAM program that applies repeated squaring to the input length and then invokes a random number generator on the result. Conversely, if the RAM program \emph{does not} receive its input length $n$ as part of its input (e.g., as described in Section~\ref{section:preliminaries}), the best achievable bound\footnote{Obtainable by starting with a constant and performing repeated squaring.} is $p_{t_0}(n) \geq 2^{-O(2^{2 t_0})}$, which only guarantees a constant failure probability. In the next section, we present explicit constructions of pure JOT-DP RAM programs in the unbounded setting where $p(n)$ vanishes as $n \to \infty$.

%% file: ram-timing-private.tex
\subsection{Pure JOT-DP RAM Programs in the Unbounded Setting}\label{subsec:ram-program-counts}
We now present a general construction for converting a JOT-DP RAM program $P$ in the upper-bounded setting into a JOT-DP RAM program $P'$ in the unbounded setting. Notably, $P'$ achieves an error probability that matches the best achievable bound (up to constant factors) within the RAM model, as established by the lower bound in Theorem~\ref{theorem:lower-bound}. Intuitively, our construction follows a similar approach to Program~\ref{program:cdl-mechanism}, but leverages the fact that the program has access to the input length $n$. The program begins by computing a coarse estimate of the dataset size, ensuring with high probability that this estimate serves as a valid upper bound on the input length $|x|$. Given this upper bound, we then apply the same technique as in Program~\ref{program:cdl-mechanism}, truncating the dataset if necessary before executing the upper-bounded JOT-DP program $P$.

Program~\ref{program:dp-count} takes as input a dataset $x \in \mathcal{D}^*$ and outputs an estimate of the dataset size using an adaptive coin-flipping process. The program initializes a biased coin with a success probability of $1/(n + k)^c$ for constants $c$ and $k$. It then repeatedly flips the coin until the first success, recording the total number of flips as the output. On the $(i+1)$th flip, the success probability is adjusted to $1/(n - i + k)^c$, and this adjustment continues until the success probability reaches a fixed value of $1/k^c$, at which point the bias is no longer updated. Thus, the PMF of the output of Program~\ref{program:dp-count} is given below.

\begin{align*}
    f_{(n, c)}(i) =
    \begin{cases} 
        \frac{1}{(n - i + k)^c}\cdot \prod_{j = 0}^{x - 1} \big(1 - \frac{1}{(n - j + k)^c}\big) & \text{if } 0 \le i \le n, \\
        \textrm{Geom}_{\frac{1}{k^c}}(i - n)\cdot\prod_{j = 0}^{n - 1} \big(1 - \frac{1}{(n - j + k)^c}\big)  & \text{if } i > n
    \end{cases}
\end{align*}
where $\textrm{Geom}_p(i)$ is the PMF of the Geometric distribution defined as:
\begin{align*}
\textrm{Geom}_p(i) =
\begin{cases} 
(1 - p)^{i-1}\cdot p & \text{if } i \in \{1, 2, 3, \dots\}, \\
0 & \text{otherwise}.
\end{cases}
\end{align*}

Importantly, the runtime of Program~\ref{program:dp-count} is fully determined by its output. Thus, if the program's output satisfies $\varepsilon$-DP, it is also true that the program will satisfy $\varepsilon$-JOT-DP.

\begin{algorithm}[ht]
\begin{flushleft}

\textbf{Input:} A dataset $x$ occupying memory locations $M[0],\dots, M[\texttt{input\_len} - 1]$. The values $c, k \ge 2$ are hardcoded constants.

\vspace{5px}
\textbf{Output:} A noisy estimate of the input length $|x|$.

\end{flushleft}
\begin{algorithmic}[1]
\STATE $\texttt{n} = \texttt{input\_len}$;
\STATE $\texttt{count} = 0$;
\STATE $\texttt{flag} = 0$;
\WHILE{$\texttt{flag} == 0$}
\STATE $\texttt{v} = \texttt{n} - \texttt{count}$; \COMMENT{Note the RAM model rounds negative numbers to $0$}
\STATE $\texttt{b} = \texttt{v} + \texttt{k}$;
\STATE $\texttt{B} = 1$;
\FOR{$\texttt{j} = 0, \dots, (c - 1)$}
\STATE $\texttt{B} = \texttt{B} \cdot \texttt{b}$; \COMMENT{Computing $\texttt{b}^c$}
\ENDFOR
\STATE $\texttt{r} = \texttt{RAND}(\texttt{B})$;
\IF{$\texttt{r} == 0$}
\STATE $\texttt{flag} = 1$;
\ELSE
\STATE $\texttt{count} = \texttt{count} + 1$;
\ENDIF
\ENDWHILE
\STATE \RETURN $\texttt{count}$;
\end{algorithmic}

\caption{A RAM Program for Approximate Timing-Private Counts}\label{program:dp-count}
\end{algorithm}

\begin{lemma}\label{lemma:pure-dp-count-private}
    For all $c \ge 2$, $k \ge 2$, the RAM program $P: \mathcal{X}\times\mathcal{E}\to\mathbb{N}\times\mathcal{E}$ described in Program~\ref{program:dp-count} is $\varepsilon$-JOT-DP in the unbounded setting where $\varepsilon = 2c\cdot\ln\big(\frac{k + 1}{k - 1}\big)$.
\end{lemma}

\begin{proof}
   Let $p_n(j) = 1/(n - j + k)^c$ and observe that 
    \begin{align*}
        p_{n+1}(j+1) = \frac{1}{(n + 1 - j - 1 + k)^c} = \frac{1}{(n - j + k)^c} = p_n(j)
    \end{align*}
    and therefore for all $y < n$
    \begin{align*}
        \prod_{j=0}^y  \Big(\frac{1 - p_n(j)}{1 - p_{n+1}(j)}\Big) = \frac{1 - p_n(y)}{1 - p_{n+1}(0)} 
        \le \frac{(n+1+k)^c}{(n+1+k)^c - 1} 
        \le \frac{k^c}{k^c - 1}
    \end{align*}
    and similarly
    \begin{align*}
        \prod_{j=0}^y  \left(\frac{1 - p_{n+1}(j)}{1 - p_{n}(j)}\right) = \frac{1 - p_{n+1}(0)}{1 - p_{n}(y)} \le \frac{(n - y + k)^c}{(n - y + k)^c - 1} \le \frac{k^c}{k^c - 1}
    \end{align*}

    We now consider the ratio of PMFs. \\
    
     \noindent\textbf{Case 1 ($y \le n$)}: 
    \begin{align*}
        \frac{f_n(y)}{f_{n+1}(y)} &= \frac{p_n(y)}{p_{n+1}(y)}\cdot \prod_{j = 0}^{y - 1} \left(\frac{1 - p_n(j)}{1 - p_{n+1}(j)}\right) \\
        &= \frac{p_n(y)}{p_{n+1}(y)}\cdot \frac{1 - p_n(y - 1)}{1 - p_{n+1}(0)} \\
        &\le \left(\frac{(n + 1 - y + k)^c}{(n - y + k)^c}\right)\cdot \frac{k^c}{k^c - 1} \\
        &\le \frac{(k + 1)^c}{k^c} \cdot \frac{k^c}{k^c - 1} \\
        &\le \frac{(k+1)^c}{k^c - 1} \\
        &\le \left(\frac{k+1}{k - 1}\right)^{2c}
    \end{align*}
    
    Similarly,
    \begin{align*}
        \frac{f_{n+1}(y)}{f_n(y)} &= \frac{p_{n+1}(y)}{p_n(y)}\cdot \prod_{j = 0}^{y - 1} \left(\frac{1 - p_{n+1}(j)}{1 - p_n(j)}\right) \\
        &= \frac{p_{n+1}(y)}{p_n(y)}\cdot \left(\frac{1 - p_{n+1}(0)}{1 - p_n(y - 1)}\right) \\
        &= \frac{(n - y + k)^c}{(n + 1 - y + k)^c}\cdot \left(\frac{k^c}{k^c - 1}\right)\\
        &\le \frac{(n - y + 1 + k)^c}{(n - y + 1 + k)^c - 1} \\
        &\le \frac{(k + 1)^c}{(k + 1)^c - 1} \\
        &\le \frac{(k + 1)^c}{k^c - 1} \\
        &\le \left(\frac{k+1}{k - 1}\right)^{2c}
    \end{align*}

    \noindent\textbf{Case 2 ($y > n$)}:
    \begin{align*}
        \frac{f_n(y)}{f_{n+1}(y)} &= \frac{\textrm{Geom}(p = \frac{1}{k^c}, y - n)}{\textrm{Geom}(p = \frac{1}{k^c}, y - n - 1)} \cdot \left(\frac{1}{1 - p_{n+1}(n)}\right) \cdot \prod_{j = 0}^{n - 1} \left(\frac{1 - p_n(j)}{1 - p_{n+1}(j)}\right) \\
        &= \left(\frac{k^c}{k^c}\right)\cdot \frac{(1 - \frac{1}{k^c})^{y - n - 1}}{(1 - \frac{1}{k^c})^{y - n - 2}} \cdot \left(\frac{(k + 1)^c}{(k + 1)^c - 1}\right) \cdot \left(\frac{1 - p_n(n - 1)}{1 - p_{n+1}(0)}\right) \\
        &\le\frac{(k + 1)^c}{(k + 1)^c - 1} \cdot \left(\frac{k^c}{k^c - 1}\right) \\
        &\le \left(\frac{k + 1}{k - 1}\right)^{2c}
    \end{align*}

    and similarly,
    \begin{align*}
    \frac{f_{n+1}(y)}{f_n(y)} &= \frac{\textrm{Geom}(p = \frac{1}{k^c}, y - n - 1)}{\textrm{Geom}(p = \frac{1}{k^c}, y - n)} \cdot \left(1 - p_{n+1}(n)\right)\cdot \prod_{j = 0}^{n - 1} \left(\frac{1 - p_{n+1}(j)}{1 - p_n(j)}\right) \\
    &\leq \frac{k^c}{k^c - 1} \cdot \left(\frac{1 - p_{n+1}(0)}{1 - p_n(n - 1)}\right) \\
    &\leq \frac{k^c}{k^c - 1}\cdot \left(\frac{k^c}{k^c - 1}\right) \\
    &\leq \left(\frac{k + 1}{k - 1}\right)^{2c}
\end{align*}

    It follows that the program is $\varepsilon$-DP for $\varepsilon = 2c\cdot \ln\big(\frac{k+1}{k - 1}\big)$. Furthermore, observe that the runtime of this program is a deterministic function of its output. Specifically, 
    \begin{align*}
        \TP{x, \env} = 5 + (\out(P(x, \env)) + 1)\cdot (7 + 2(c-1))
    \end{align*}

    Since there exists a deterministic function $f$ such that $\TP{x, \env} = f(\out(P(x,\env)))$, if $\out(P(x,\env))$ is $\varepsilon$-DP, then by post-processing, $(\out(P(x, \env)), \TP{x,\env})$ must also satisfy $\varepsilon$-DP.
    Thus, the claim follows. 
\end{proof}

We now describe how to set $k$ appropriately. By Lemma~\ref{lemma:pure-dp-count-private}, we have
\begin{align*}
\varepsilon \ge 2c \cdot \ln\!\left(\frac{k+1}{k-1}\right) = 2c\cdot \ln\!\left(1 + \frac{2}{k-1}\right)
\end{align*}

When $k$ is large, we can use the approximation that $\ln(1+x) \approx x$ for small $x$, and therefore
\begin{align*}
\varepsilon &= 2c \cdot \ln\!\left(1 + \frac{2}{k-1}\right) \\
           &\approx 2c \cdot \frac{2}{k-1} \\
           &= \frac{4c}{k-1}
\end{align*}
Rearranging the inequality, it suffices to choose $k = O\!\left(c/\varepsilon\right)$ and then run Program~\ref{program:dp-count} to obtain a JOT-DP estimate of the input length of a RAM program. We now demonstrate that this estimate is sufficiently accurate to derive an \emph{upper bound} on the input length, where the probability that the upper bound is less than the true length of the input decays inverse polynomially in $n$.

\begin{lemma}\label{lemma:accuracy-dp-count}
For all $c \ge 2, k \ge 2$, the RAM program $P: \mathcal{X}\times\mathcal{E}\to\mathbb{N}\times\mathcal{E}$ described in Program~\ref{program:dp-count} outputs an estimate $\hat{y} = \out(P(x,\env))$ of $|x| = n$ that satisfies 
$$\Pr\Big[\hat{y} < \frac{n}{2}\Big] \leq O\Big(\frac{1}{n^{c - 1}}\Big)$$
\end{lemma}

\begin{proof}
Let $n = |x|$ and $\hat{y} = \out(P(x, \env))$ be the output of Program~\ref{program:dp-count}. We have that
    \begin{align*}
        \Pr[\hat{y} < \frac{n}{2}] &= \sum_{i = 1}^{\frac{n}{2}} \left(p_n(i) \prod_{j = 0}^{i - 1} (1 - p_n(j))\right) \\
        &\le \sum_{i = 1}^{\frac{n}{2}} p_n(i) \hspace{1cm} \text{(union bound)} \\
        &\le \frac{n}{2}\cdot p_n\Big(\frac{n}{2}\Big) \\
        &= \frac{\frac{n}{2}}{(\frac{n}{2} + k)^c} \\
        &\le O\Big(\frac{1}{n^{c-1}}\Big) \\
    \end{align*}

    and the claim follows.
\end{proof}

By Lemma~\ref{lemma:accuracy-dp-count}, we can set \( m = 2\cdot\hat{y} \), ensuring that with probability \( O(1/n^{c-1}) \), we have \( m \geq |x| \). We can then apply the same technique as in Section~\ref{sec:pure-timing-privacy} and compose Program~\ref{program:dp-count} with an arbitrary RAM program $P$ that is JOT-DP in the upper-bounded setting. We describe the general construction in Program~\ref{program:optimal-ram}.

\begin{algorithm}[t]
\begin{flushleft}

\textbf{Input:} A dataset $x$, and a privacy parameter $\varepsilon_1 > 0$. The program $P$ is hardcoded into line 4

\vspace{5px}
\textbf{Output:} A noisy estimate of the input length $|x|$.

\end{flushleft}
\begin{algorithmic}[1]
\STATE Set $\hat{n}$ to the output of  Program~\ref{program:dp-count} with constants $k$ and $c$ such that $k = O(c/\varepsilon')$;
\STATE $\hat{n} = 2\cdot \hat{n}$;
\STATE $\hat{x} = \texttt{Truncate}(x, \hat{n})$; \COMMENT{returns the first $\hat{n}$ entries of $x$ if $|x| > \hat{n}$}
\STATE $\textrm{run $\varepsilon_2$-JOT-DP program $P^{\hat{n}}(\hat{x})$ and return the result}$ \COMMENT{$P^{\hat{n}}$ bounds inputs to length $\hat{n}$}
\end{algorithmic}

\caption{JOT-DP RAM Program for the Unbounded Setting}\label{program:optimal-ram}
\end{algorithm}

    \begin{lemma}\label{lemma:pure-dp-count-runtime-linear}
        For all $c \geq 2$, $k \geq 2$, the RAM program $P: \mathcal{X} \times \mathcal{E} \to \mathbb{N} \times \mathcal{E}$ described in Program~\ref{program:dp-count} runs in time $O(n)$ with probability at least $1 - (1 - \frac{1}{k^c})^n$, where $n = |x|$.
    \end{lemma}
    
    \begin{proof}
        The runtime of Program~\ref{program:dp-count} is a deterministic function of its output. Specifically,

        \begin{align*}
            \TP{x, \env} &= 5 + (\out(P(x, \env)) + 1) \cdot (6 + 2(c-1)) \\
            &= O(\out(P(x, \env)))
        \end{align*}

        Thus, it suffices to show that $\Pr[\out(P(x,\env)) \leq 2n]$ holds with high probability. Observe that:

        \begin{align*}
            \Pr[\out(P(x, \env)) > 2n] &\leq \textrm{Geom}\Big(p = \frac{1}{k^c}, n\Big) \\
            &\leq \Big(1 - \frac{1}{k^c}\Big)^n
        \end{align*}
    \end{proof}

\begin{theorem}[Pure JOT-DP RAM Programs]\label{theorem:dp-sum-unbounded}
           For all $\varepsilon > 0$, $\varepsilon' > \varepsilon$, and $\varepsilon$-DP RAM programs $P:\mathcal{X}\times\mathcal{E}\to\mathcal{Y}\times\mathcal{E}$ in the upper-bounded setting, there exists a $\varepsilon'$-JOT-DP RAM program $P':\mathcal{X}\times\mathcal{E}\to\mathbb{N}\times\mathcal{E}$ for the unbounded setting satisfying $\forall x \in\mathcal{X}$ of length $n$, $\forall$ input-compatible $\env\in\mathcal{E}$, $\forall c \ge 2$:
    \begin{align*}
        \Big\|\out(P(x, \env)) - \out(P'(x, \env))\Big\|_{TV} < O\left(\frac{1}{n^c}\right)
    \end{align*}
    Furthermore, the mechanism $P'(x, \env)$ is an explicit algorithm that makes one oracle call to $P$ on a truncation of $x$ along with an additional computation that takes time $O(|x|)$ with high probability.
\end{theorem}

\begin{proof}
    The proof follows a similar structure to that of Theorem~\ref{theorem:constructive}. In line 1 of Program~\ref{program:optimal-ram}, a noisy estimate $\hat{n}$ of $n = |x|$ is obtained by executing Program~\ref{program:dp-count}. By Lemma~\ref{lemma:pure-dp-count-private}, Program~\ref{program:dp-count} satisfies $\varepsilon_1$-JOT-DP in the unbounded setting. Furthermore, lines 2-3 can be executed in a fixed number of instructions. Thus, we represent lines 1-3 of Program~\ref{program:dp-count} as the $\varepsilon_1$-JOT-DP program $P_{\texttt{(line 3)}}$. 

    Next, we consider the composition of $P_{\texttt{(line 3)}}$ with the $\varepsilon_2$-JOT-DP program $P$. By sequential composition, it follows that Program~\ref{program:dp-count} is $\varepsilon'$-JOT-DP in the unbounded setting, where $\varepsilon' = \varepsilon_1 + \varepsilon_2$. What remains to be shown are the accuracy and runtime guarantees of Program~\ref{program:optimal-ram}. 
    
    By Lemma~\ref{lemma:pure-dp-count-runtime-linear}, we have that, with high probability, Program~\ref{program:dp-count} runs in time $O(n)$ before making a single oracle call to $P$ on a truncated version of $x$. Observe that, conditioned on $\hat{n} > \frac{n}{2}$ in line 1, we have $\hat{x} = x$ after executing line 3. Thus, Program~\ref{program:optimal-ram} returns the output of the $\varepsilon$-JOT-DP program $P$ on input $x$. By Lemma~\ref{lemma:accuracy-dp-count}, Program~\ref{program:dp-count} in line 1 returns an estimate $\hat{n}$ such that 

    \begin{align*}
        \Pr\Big[\hat{n} < \frac{n}{2}\Big] = O\left(\frac{1}{n^{c-1}}\right)
    \end{align*}

    Therefore, after executing line 2 so that $\hat{n} = 2\cdot \hat{n}$, it follows that $\hat{n} > |x|$ with probability at least $1 - O(\frac{1}{n^{c-1}})$. Conditioning on this case, $\hat{x} = x$ after executing line 3, and therefore Program~\ref{program:dp-count} will return the output of $P^{\hat{n}}$ on input $x$, where $P^{\hat{n}}$ accepts inputs of length at most $\hat{n}$ and satisfies $\hat{n} > |x|$. 
\end{proof}

\begin{corollary}[Pure JOT-DP Laplace Mechanism]\label{corollary:jot-dp-laplace}
    For all $\varepsilon > 0$, and all $c > 0$, there exists an $\varepsilon$-JOT-DP RAM program $P:\mathcal{X} \times \mathcal{E} \to \mathcal{Y} \times \mathcal{E}$ for releasing sums in the unbounded setting such that for all datasets $x \in \mathcal{X}$:
    \begin{align*}
        \Pr\left[\Big|\out(P(x, \env)) - \sum x_i\Big| \geq \frac{C\cdot \ln(n)}{\varepsilon}\right] < O\left(\frac{1}{n^c}\right)
    \end{align*}
    where $n = |x|$ is the size of the dataset, and $C > 0$ is a universal constant.
\end{corollary}
\begin{proof}
Fix $c \ge 2$. Let $M$ be the Censored Discrete Laplace mechanism that adds noise drawn from a Discrete Laplace distribution and clips the output to lie within the bounds $\ell = 0$ and $u = \Delta \cdot U$, where $U$ is an upper bound on the input length and each $x_i \in [0, \Delta]$.  
By Lemma~\ref{lemma:censored-laplace-jot-dp}, there exists an $\varepsilon'$-JOT-DP RAM program 
$P_{\text{Lap}}: \mathcal{X} \times \mathcal{E} \to \mathcal{Y} \times \mathcal{E}$ that implements $M$, and furthermore $P_{\text{Lap}}$ is a $\varepsilon'$-JOT-DP RAM program for the upper-bounded setting since it accepts inputs of length at most $U$. Let $Z$ denote the output of the censored Discrete Laplace program $P_{\text{Lap}}$. Because $Z \sim \texttt{CensoredDiscreteLaplace}(\mu = \sum x_i, s=\Delta/\varepsilon', \ell = 0, u=\Delta\cdot U)$, we have
\begin{align*}
    \Pr[|Z - \mu| \ge t] \le \exp\left(-\frac{\varepsilon'\cdot t}{\Delta}\right)
\end{align*}

Applying Theorem~\ref{theorem:dp-sum-unbounded} with $\varepsilon > \varepsilon'$
gives an $\varepsilon$-JOT-DP RAM program
$P:\mathcal{X}\times\mathcal{E}\to\mathcal{Y}\times\mathcal{E}$ such that, for every $x$ of length $n$ and every input-compatible $\env$
\begin{align*}
        \bigl\|\out(P_{\text{Lap}}(x,\env))-\out(P(x,\env))\bigr\|_{TV}=O\!\bigl(n^{-c}\bigr)
\end{align*}

Thus, by a union bound: 
\begin{align*}
        \Pr\!\Bigl[\bigl|\out(P(x,\env))-\sum_{i=1}^{n}x_i\bigr|\ge t\Bigr]
        \le \exp\left(-\frac{\varepsilon'\cdot t}{\Delta}\right) + O\bigl(n^{-c}\bigr)
\end{align*}
We choose $t = \Delta\cdot c\ln n/\varepsilon'$ so that 
\begin{align*}
        \Pr\!\Bigl[\bigl|\out(P(x,\env))-\sum_{i=1}^{n}x_i\bigr|\ge t\Bigr]
        \le \frac{1}{n^c} + O\bigl(n^{-c}\bigr) = O\bigl(n^{-c}\bigr)
\end{align*}

Thus, setting $C=\Delta\cdot c \cdot \varepsilon / \varepsilon' $ and substituting $t=C\ln n/\varepsilon$:
\begin{align*}
        \Pr\!\Bigl[\bigl|\out(P(x,\env))-\sum_{i=1}^{n}x_i\bigr|
                \ge\frac{C\ln n}{\varepsilon}\Bigr]
        <O\!\bigl(n^{-c}\bigr)
\end{align*}
\end{proof}

%% file: appendix.tex
\section{Appendix}
We show that programs that attempt to avoid \emph{all} information leakage through their runtime (i.e., $\TP{x, \env} \equiv \TP{x', \env'}$ for all $x, x' \in \mathcal{D}^*$ and $\env, \env' \in \mathcal{E}$) will experience some loss in utility. In particular, we show that programs for computing means will exhibit a constant additive error in their output even as the length of the input $n\to\infty$. This result suggests that some portion of the program's privacy budget must be allocated to privatizing the program's runtime.

\begin{lemma}\label{lemma:perfect-timing-privacy-implies-bad-accuracy}
    Let $P: \mathcal{X}\times\mathcal{E}\to \mathcal{Y}\times\mathcal{E}$ be a $\varepsilon$-DP RAM program for computing the mean $\frac{1}{|x|}\sum_{i = 1}x_i$ such that $\TP{x, \env} \equiv \TP{x', \env'}$ for all $x, x' \in \{0, 1\}^*$ and input-compatible $\env, \env' \in \mathcal{E}$. Then for every $0 < \beta < 1$, there exists a dataset $x$ and a constant $\alpha$ such that 
    \begin{align*}
        \Pr\left[\left|\out(P(x, \env)) - \frac{1}{|x|}\sum x_i\right| > \alpha\right] > \beta
    \end{align*}
\end{lemma}
\begin{proof}
    Let $F$ be the cumulative distribution function of the runtime random variable $\TP{x, \env}$. We pick $p > 0$ and let $t = F^{-1}(1 - p)$. Then $\Pr[\TP{x, \env} > t] = p$. When $\TP{x, \env} \le t$, the program can read at most $t$ entries of the input. By lower bounds for samplers~\cite{canetti1995lower}, for every $0 < \beta < 1$, there exists an input $x$ such that, conditioned on reading at most $t$ locations of $x$, the algorithm fails to output an $\alpha$-accurate estimate $\hat{\mu}$ of $\mu = \frac{1}{|x|}\sum x_i$ with probability at least $\beta$ for 
    \begin{align*}
        \alpha = \Omega\left(\sqrt{\frac{\log (1/\beta)}{t}}\right)
    \end{align*}

    Thus, without conditioning, the program fails to output an $\alpha$-accurate estimate of $\mu$ with probability at least $\beta - p$. We take $\beta = 2p$ so that $t$ is a fixed constant and hence so is $\alpha$. 
\end{proof}

\begin{lemma}[No Fully-Dyadic Censored Discrete Laplace]\label{lemma:no-dyadic-dl}
Fix integers $\ell<\mu < u$ and a scale $s>0$.  
Let $Z$ be the Discrete Laplace random variable with shift $\mu\in\mathbb{Z}$,  
censored to the support $\{\ell, \ell + 1, \dots, \mu, \mu + 1, \dots, u\}$:
\begin{align*}
\Pr[Z = k] = 
\begin{cases}
F(\ell \mid \mu, s) & \text{if } k = \ell \\[6pt]
p(k \mid \mu, s) & \text{if } \ell < k < u \\[6pt]
1 - F(u - 1 \mid \mu, s) & \text{if } k = u \\[6pt]
0 & \text{otherwise}
\end{cases}
\end{align*}
where
\begin{align*}
p(k \mid \mu, s) &= \frac{e^{1/s} - 1}{e^{1/s} + 1} \cdot e^{-|k - \mu| / s} \\[6pt]
F(x \mid \mu, s) &=
\begin{cases}
\frac{e^{1/s}}{e^{1/s} + 1} \cdot e^{-(\mu - x)/s} & \text{if } x \leq \mu \\[6pt]
1 - \frac{1}{e^{1/s} + 1} \cdot e^{-(x - \mu)/s} & \text{if } x > \mu
\end{cases}
\end{align*}
If the support contains at least four points, then there exists a $k \in \{\ell, \ell + 1, \dots, \mu, \mu + 1, \dots, u\}$ such that:
\begin{align*}
\Pr[Z=k]\notin\Bigl\{\tfrac{t}{2^{N}}:t,N\in\mathbb{N}\Bigr\}
\end{align*}
That is, there is at least one output mass that is non-dyadic.
\end{lemma}

\begin{proof}
Let $q=e^{-1/s}\in(0,1)$ and $c=\dfrac{1-q}{1+q}$.
For every interior index $\ell<k<u$ we have
\[
\Pr[Z=k]=c\cdot q^{|k-\mu|}
\]

\medskip\noindent
\textbf{Case 1: $q$ is dyadic.}
Write $q=m/2^N$ with $1\le m<2^N$ odd.  Then
\[
c=\frac{1-q}{1+q}=\frac{2^N-m}{2^N+m}
\]
whose denominator $2^N+m$ is not a power of two, so $c$ is
non-dyadic.  Because $\ell<\mu<u$, it follows that
$\Pr[Z=\mu]=c$ is a non-dyadic mass.

\medskip\noindent
\textbf{Case 2: $q$ is \emph{not} dyadic.}
Write $q = r/d$ in lowest terms.  
Since $q$ is not dyadic, the denominator $d$ is not a 
power of two, so it contains at least one odd prime factor. Because the support has $\ge4$ points, either 
$\ell < \mu + 1 < u$ or $\ell < \mu - 1 < u$. Assume $\ell < \mu + 1 < u$ (the other side
is symmetric).  Then
\begin{align*}
\Pr[Z=\mu+1]=c\cdot q
           =\frac{d-r}{d+r}\cdot\frac{r}{d}
\end{align*}

Let $j$ be the odd prime dividing $d$.
Because $\gcd(r,d)=1$, $j$ does not divide $r$. Consequently $j$ does not divide $(d - r)$ or $(d+r)$.  
Thus the denominator $d\cdot (d+r)$ contains at least one factor $j$,
whereas the numerator $r\cdot (d - r)$ contains none.
After cancelling the greatest common divisor, a power of $j$ remains
in the denominator, so the reduced fraction is not of the form
$t/2^{N}$.  Hence $\Pr[Z=\mu+1]$ is non-dyadic.

\end{proof}

\begin{lemma}[Finite-Coin Sampler for Censored Dyadic Symmetric Geometric]
\label{lemma:cdtg-finite}
Let integers $\ell < \mu < u$ and $p \in (0,1)$ (dyadic rational) be given.  
Draw one unbiased coin for a sign $S\in\{-1,+1\}$ and draw $G\sim\operatorname{Geom}(p)$.  
Define a Dyadic Symmetric Geometric random variable as:
\[
Z=
\begin{cases}
\mu - G & \text{if } S = -1\\[4pt]
\mu + 1 + G & \text{if } S = +1
\end{cases}
\]
Then a \emph{Censored} Dyadic Symmetric Geometric random variable  
\[
Y = \max\{\ell,\min\{Z,u\}\}
\]

is exactly sampleable with a finite (constant) number of unbiased coin flips under the $\text{RAM}_{\text{BDDNT}}$ model.
\end{lemma}

\begin{proof}
Flip one unbiased coin to choose a sign $S \in \{-1,+1\}$ and draw $G \sim \operatorname{Geom}(p)\!\big|_{0}^{m}$, the geometric distribution with parameter $p = 2^{-k}$ clamped to $\{0,1,\dots,m\}$ where $m = \max\{\mu - \ell, u - \mu\}$. Each Bernoulli$(p)$ trial in the geometric sampler can be implemented using exactly $k$ unbiased coins, so the clamped geometric draw uses at most $k \cdot (m + 1)$ coin flips. Together with the one coin for $S$, the entire sampling procedure uses at most $1 + k \cdot (m + 1)$ unbiased coins. Once $S$ and $G$ are drawn, define $Y = \mu - G$ if $S = -1$ and $Y = \mu + 1 + G$ if $S = +1$. The value $Y$ is a sample from the clamped Dyadic Symmetric Geometric distribution centered at $\mu$ and clamped to $[\ell, u]$ as desired. Therefore, $Y$ is sampleable with a finite (constant) number of unbiased coin flips under the $\text{RAM}_{\text{BDDNT}}$ model.
\end{proof}

\begin{lemma}[Censored Dyadic Symmetric Geometric Mechanism is DP]
\label{lemma:cdtg-privacy}
Fix integers $\ell < \mu < u$ and let $\mu' = \mu + 1$.  
Let $p \in (0,1)$ be any dyadic rational.  
Let $Z \sim \operatorname{DTG}_{\mu,p}$ and $Z' \sim \operatorname{DSG}_{\mu',p}$ be samples from the (unclamped) Dyadic Symmetric Geometric distribution centered at $\mu$ and $\mu'$, respectively.  
Let $Y = \max\{\ell, \min\{Z, u\}\}$ and $Y' = \max\{\ell, \min\{Z', u\}\}$.  
Then for every $y \in \mathbb{Z} \cap [\ell,u]$ we have
\begin{align*}
\frac{\Pr[Y = y]}{\Pr[Y' = y]} \le \frac{1}{1 - p}
\quad \text{and} \quad
\frac{\Pr[Y' = y]}{\Pr[Y = y]} \le \frac{1}{1 - p}
\end{align*}
so the mechanism is $\varepsilon$-differentially private with
\begin{align*}
\varepsilon = \ln\!\left( \tfrac{1}{1 - p} \right)
\end{align*}
\end{lemma}

\begin{proof}
We analyze the \emph{unclamped} Dyadic Symmetric Geometric random variables $Z$ and $Z'$.  
Generate $Z$ by flipping one unbiased coin for a sign $S \in \{-1,+1\}$ and drawing $G \sim \operatorname{Geom}(p)$, setting $Z = \mu - G$ if $S = -1$ and $Z = \mu + 1 + G$ if $S = +1$. Generate $Z'$ in the same way but with center $\mu' = \mu + 1$.  For every $z \in \mathbb{Z}$ the shift of the center changes the probability mass by \emph{at most} a single factor of $q = 1 - p$, so
\begin{align*}
\frac{\Pr[Z = z]}{\Pr[Z' = z]} \le q^{-1}
\quad \text{and} \quad
\frac{\Pr[Z' = z]}{\Pr[Z = z]} \le q^{-1}
\end{align*}
Since the ratio for $Z$ and $Z'$ is bounded by $1/(1-p)$, the random variables $Z$ and $Z'$ are $\varepsilon$–differentially private with
\begin{align*}
\varepsilon = \ln\!\left( \tfrac{1}{1 - p} \right)
\end{align*}
The outputs $Y = \max\{\ell, \min\{Z, u\}\}$ and $Y' = \max\{\ell, \min\{Z', u\}\}$ are obtained from $Z$ and $Z'$ by deterministic clamping, which is a post-processing operation. Consequently the same bound holds for every $y \in \mathbb{Z} \cap [\ell,u]$ and the clamped mechanism is also $\varepsilon$–DP.

\end{proof}

\begin{theorem}[Pure JOT-DP $\text{RAM}_{\text{BDDNT}}$ Programs in the Unbounded Setting]
\label{theorem:constructive-bddnt}
For all $0 < \beta < 1$, $\varepsilon > 0$, $\varepsilon' > \varepsilon$, and $\varepsilon$-JOT-DP $\text{RAM}_{\text{BDDNT}}$ programs $P:\mathcal{X}\times\mathcal{E}\to\mathcal{Y}\times\mathcal{E}$ in the \emph{upper}-bounded setting, there exists a $\varepsilon'$-JOT-DP $\text{RAM}_{\text{BDDNT}}$ program $P':\mathcal{X}\times\mathcal{E}\to\mathbb{N}\times\mathcal{E}$ for the unbounded setting such that
    \begin{align*}
        \Big\|\out(P(x, \env)) - \out(P'(x,\env)) \Big\|_{TV} < \beta
    \end{align*}
   Furthermore, the mechanism $P'(x, \env)$ is simply an explicit algorithm that makes one oracle call to $P$ on a truncation of $x$ along with an additional computation that takes time $O(|x|)$ with probability at least $1 - \beta - e^{-\Omega(|x|)}$.
\end{theorem}

\begin{proof}
By replacing the call to $\texttt{CensoredDiscreteLaplace}$ in Program~\ref{program:cdl-mechanism} with the Censored Dyadic Symmetric Geometric Mechanism $\texttt{CDSG}$, we obtain an equivalent result for Theorem~\ref{theorem:constructive} in the $\text{RAM}_{\text{BDDNT}}$ model. This follows directly from Lemma~\ref{lemma:cdtg-finite} and Lemma~\ref{lemma:cdtg-privacy} which give the needed $\varepsilon_i$-DP mechanism that executes in constant-time during each iteration of the program's loop. The rest of the proof follows identically to the proof of Theorem~\ref{theorem:constructive}. In particular, during each iteration of the loop, when $m_i \ge (2/\varepsilon_i)\cdot \ln(1/\beta_i)$ we can round $p = 1 - e^{-\varepsilon_i}$ \emph{down} to the nearest dyadic rational. Then when $m_i < |x|$, we have that for $\hat{n}_i = \texttt{CDSG}(\mu = m_i, p = 1 - e^{-\varepsilon_i}, \ell = 0, u = m_i)$:
\begin{align*}
    \Pr\bigl[\hat{n}_i < \tfrac{m_i}{2}\bigr] = \frac{(1 - p)^{\tfrac{m_i}{2}}}{2} \le e^{-m_i \cdot \varepsilon_i / 2} \le \beta_i
\end{align*}

as desired. Furthermore, if we let iteration $j$ be the first iteration when $m_j > 4\cdot |x|$, then for $\hat{n}_j = \texttt{CDSG}(\mu = |x|, p = 1 - e^{-\varepsilon_j}, \ell = 0, u = m_j)$:
\begin{align*}
    \Pr\bigl[\hat{n}_j > \tfrac{m_j}{2}\bigr] = \frac{(1 - p)^{\tfrac{m_j}{2} - |x|}}{2}  \le e^{-\varepsilon_j\cdot |x|}\le e^{-\Omega(|x|)}
\end{align*}

Thus, with high probability the program halts on iteration $j$, and therefore the runtime analysis follows exactly to that of Theorem~\ref{theorem:constructive}.
\end{proof}